\documentclass[sd,amsmath,amssymb,reprint,graphicx]{revtex4-1}

\usepackage{graphicx}
\usepackage{dcolumn}
\usepackage{bm}
\usepackage{siunitx}

\draft 

\begin{document}



\title{Waveguide-loaded silica fibers for coupling to high-index micro-resonators} 



\author{P. Latawiec}
\author{M. J. Burek}
\author{V. Venkataraman}
\author{M. Lon\v{c}ar}
\affiliation{John A. Paulson School of Engineering and Applied Science, Harvard University}


\date{\today}

\begin{abstract}
Tapered silica fibers are often used to rapidly probe the optical properties of micro-resonators. However, their low refractive index precludes phase-matching when coupling to high-index micro-resonators, reducing efficiency. Here we demonstrate efficient optical coupling from tapered fibers to high-index micro-resonators by loading the fibers with an ancillary adiabatic waveguide-coupler fabricated via angled-etching. We demonstrate greatly enhanced coupling to a silicon multimode micro-resonator when compared to coupling via the bare fiber only. Signatures of resonator optical bistability are observed at high powers. This scheme can be applied to resonators of any size and material, increasing the functional scope of fiber coupling.
\end{abstract}


\maketitle 


Efficient coupling of light to and from an integrated chip is essential for many applications in nonlinear optics\cite{Eggleton2011, Cardenas2013, Lu2014a, Hausmann2014, Xiong2012, Choy2012, Lake2015}, optomechanics\cite{Mitchell2013, Khanaliloo}, and quantum optics\cite{Faraon2010, Hausmann2013a}. For these reasons, a wide range of techniques with which to deliver light on-chip have been developed, such as end-fire coupling\cite{Brenner1994}, planar gratings\cite{Taillaert2002}, or optical fiber coupling to dispersion-engineered photonic crystal waveguides\cite{Barclay2004a, Barclay2004}. However, many of these applications are based on materials where full integration with couplers is difficult or rapid testing is desired. In this case, silica fiber tapers, where an optical fiber is pulled to have a thin coupling region with thickness on the order of the wavelength, have formed the foundation for many optical experiments\cite{Knight1997, Cai2000} in diverse environments\cite{Riviere2013}. In particular, this platform has shown ideal coupling to silica resonators\cite{Cai2000}, with the ability to tune from under- to over-coupled based on taper positioning.

For high-index micro-resonators, the refractive index contrast results in a large propagation constant mismatch between the resonator and the tapered silica fiber, limiting the coupling efficiency into the resonator. This problem is exacerbated in larger, multimodal resonators where the fundamental mode is confined mostly in the material. Existing schemes, such as prism coupling\cite{Ilchenko2013}, or separate on-chip waveguides brought near to the resonator of interest\cite{Broaddus2009a}, are bulky and do not allow for rapid testing of multiple components. In this letter we demonstrate a method to improve upon the index matching by introducing an ancillary, support waveguide attached directly to the optical fiber. Using this platform, we show efficient coupling to the fundamental modes of a  multimodal silicon micro-resonator. In particular, we fabricate a tapered, free-standing, angle-etched silicon waveguide which we then detach and affix to the tapered section of an optical fiber. This achieves adiabatic mode conversion between the optical fiber and the silicon waveguide, effectively changing the propagation constant of the input light to that of silicon, matching it to the resonator.

Energy transfer between waveguides can be modeled via coupled mode theory\cite{Yariv1973}. For two waveguides in close proximity with propagation constants $\beta_1$ and $\beta_2$, field amplitudes $A_1$ and $A_2$, and mutual coupling $\kappa$, the coupled mode equations may be written as

\begin{equation}
\begin{cases}
\frac{dA_1}{dz} = i \kappa A_2e^{i(\beta_2-\beta_1)z} \\
\frac{dA_2}{dz} = i \kappa^* A_1e^{-i(\beta_2-\beta_1)z}
\end{cases}
\end{equation}

where $z$ is the length dimension along the direction of propagation. These equations can be solved analytically. In particular, if light is only injected into one waveguide, then the power in the second waveguide as a function of distance is given by

\begin{equation}
P(z) = \frac{1}{1+(\delta/\kappa)^2} \sin^2(\sqrt{\kappa^2+\delta^2}z)
\end{equation}

where we define $\delta = (\beta_2-\beta_1)/2$. Under this formulation, the maximum power coupled into the second waveguide cannot exceed $P_\mathrm{max} = 1/(1+(\delta/\kappa)^2)$. The corresponding length such that $P(L_c) = P_\mathrm{max}$ is given as

\begin{equation}
\label{Eq:Lc}
L_c = \frac{\pi}{2\sqrt{\kappa^2+\delta^2}}
\end{equation}

which indirectly depends on wavelength through $\delta$ and $\kappa$. In particular, for a fixed coupling constant $\kappa$, the power coupled in ($P_\mathrm{max}$) decreases rapidly as a function of increasing effective index mismatch ($\delta$). This underlies the difficulty of coupling to high-index micro-resonators from optical fibers.

In contrast to coupled mode theory, adiabatic coupling requires the breaking of translational symmetry in order to transfer optical energy across waveguides.  This technique has been used to couple light out of optical fibers and into target waveguides made of silicon\cite{Groblacher2013}, silicon nitride\cite{Thompson2013, Tiecke2015} and diamond\cite{Patel2015}. Here, we adiabatically transfer the mode from an optical fiber into such a waveguide (referred hereafter as the ``loader waveguide"), using the converted mode to efficiently couple to an index-matched micro-resonator.

The loader waveguide and resonator were fabricated via angled-etching within a Faraday cage\cite{Burek2012b, Cho}. After a resist mask is defined, the pattern is first etched vertically. In a second step, a Faraday cage is placed around the sample, directing the incident etching ions to the substrate at an angle defined primarily by the cage geometry\cite{Burek2012b}. In order to leave the structure suspended, the etch is timed to end before the structure is completely undercut\cite{Burek2014}. The final etch profile shows a three-dimensional taper since the etch depth is defined via the width from the etch angle. This can yield adiabatic coupling over shorter intervals compared to thin film platforms\cite{Patel2015, Broaddus2009a}. Because the operating principle of angled-etching is agnostic to the etch chemistry, it has been used successfully in a number of materials, including diamond\cite{Burek2012b, Burek2013, Burek2014, Bayn2014}, quartz\cite{Latawiec}, and silicon\cite{Cho, Latawiec}. 

\begin{figure}
\includegraphics{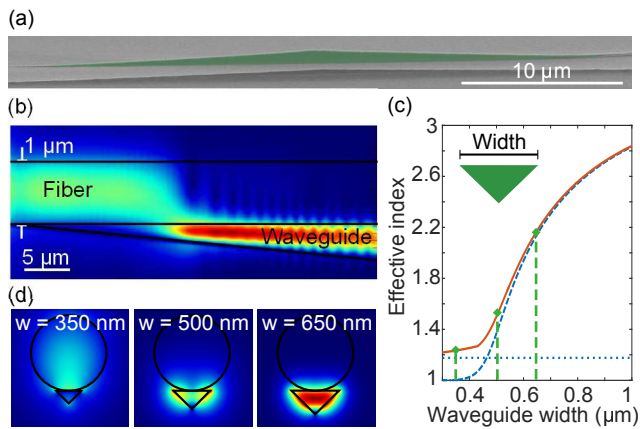}
\caption{\label{Fig1}Silicon adiabatic coupler loaded to tapered fiber. (a) Scanning electron micrograph (SEM) of a silicon loader waveguide fabricated via Faraday cage angled-etching. The device (green) tapers down from a nominal width of ${1}\mathrm{\mu m}$ in the center to points at the ends over a length of ${20}\mathrm{\mu m}$. The supporting fin is etched through, collapsing the device onto the substrate (b) Finite-difference time-domain (FDTD) simulation of a silicon adiabatic coupler loaded to a tapered silica fiber with diameter ${1}\mathrm{\mu m}$. The mode originally stays in the silica fiber before being drawn into the waveguide. The waveguide is shown coupling over ${40}\mathrm{\mu m}$. The multimodal nature of the thicker end of the silicon waveguide visibly manifests itself as interference fringes in the normalized electric field. (c) COMSOL simulations of the effective index of the tapered fiber (blue, dotted), silicon coupler device (blue, dashed), and supermode (red, solid) as the width of the silicon coupler is increased from 300nm to ${1}\mathrm{\mu m}$. The dashed green lines correspond to the cross-sections shown in (d). The waveguide's cross-section is an isosceles triangle with equal angles of 30$^\circ$. (d) Supermodes obtained at different cross-sectional waveguide widths.}
\end{figure}

\begin{figure}
\includegraphics{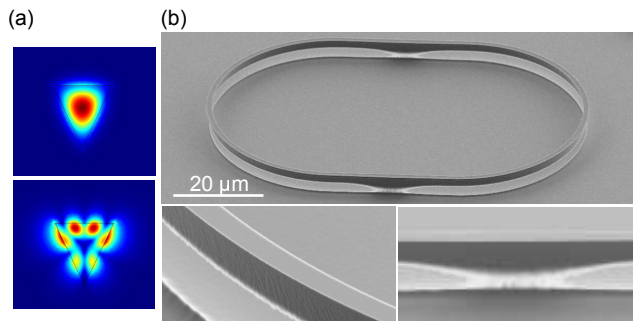}
\caption{\label{Fig2}Studied silicon micro-resonator with nominal width ${1}\mathrm{\mu m}$. (a) Simulated modal profiles at $\lambda_0 = 1.52 \mathrm{\mu m}$ for the fundamental (top, $n_{\textrm{eff}} = 3.19, \ n_g = 3.85$) and higher-order (bottom, $n_{\textrm{eff}} = 1.52$) modes using an extracted etch angle of $65^\circ$. (b) SEM image of silicon micro-resonator (top) fabricated with angled-etching. The bending radius is ${25}\mathrm{\mu m}$ while the straight region is ${25}\mathrm{\mu m}$. A hydrogen silsesquioxane (HSQ) mask is defined via electron-beam lithography and the structure is etched in a SF$_6$/C$_4$F$_8$ chemistry. Slight waveguide roughness (bottom left) is visible on the SEM. The structure is supported above the substrate by silicon fins (bottom right). This is done by timing the etch so that material remains under the widened support sections.}
\end{figure}

Once sample etching is complete, the resist is removed and the loader waveguide is manually detached from the substrate. Thereafter, it is transferred to a tapered optical fiber (diameter $\sim{1}{\mathrm{\mu m}}$) by bringing the fiber into contact as it lies on the substrate. As the waveguide is completely detached from the substrate, the fiber's attractive forces are sufficient to pull the device off. Afterwards, the loader waveguide is manipulated against features on the sample until it lies parallel to the fiber. During the loading process no visible damage is done to the fiber, although the fiber transmission reduces to $\sim 0.92 \%$. As the loading process is not done in a controlled environment, large scattering centers such as dirt or dust on the silicon chip can be picked up by the fiber. Alternate methods of loading\cite{Patel2015, Tiecke2015} can increase the overall transmission of this design.

\begin{figure*}
\includegraphics{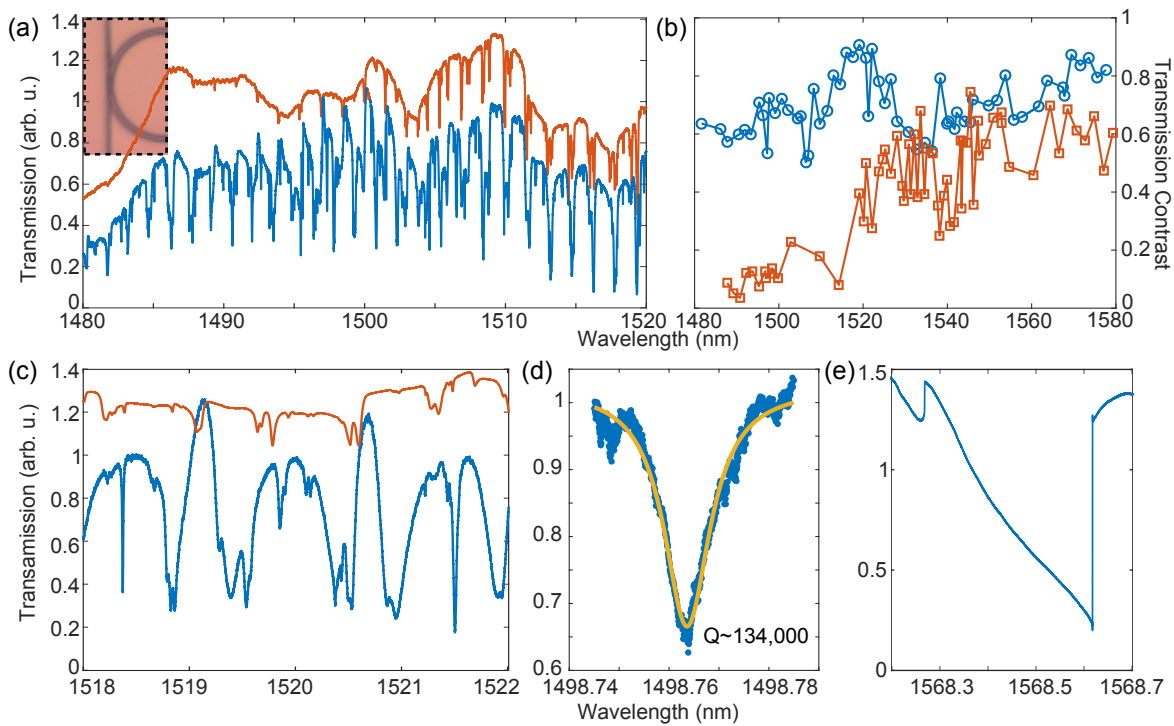}
\caption{\label{Fig3}Silicon micro-resonator spectra with bare fiber (red) and loaded fiber (blue) (a) Transmission measurements of silicon micro-resonator (inset, optical image) from ${1480}\mathrm{nm}$ to ${1520}\mathrm{nm}$. The bare fiber result is shifted from the loaded fiber result for clarity. In the bare fiber case, the transmission dips are not consistent across the entire spectrum, petering off at shorter wavelengths. In contrast, the loaded fiber shows a consistent coupling for all modes in the same family, with transmission dips of $\sim 40-60\%$. (b) Extracted transmission contrast for highly-coupled resonances as a function of wavelength under bare (red) and waveguide-loaded (blue) coupling. (c) Optimized coupling to high-quality factor (Q) resonances at ${\sim 1518.4}\mathrm{nm}$ and ${\sim 1521.5}\mathrm{nm}$. The measured Qs were $\sim130,000$ and $\sim40,000$, with transmission dips of 60\% and 80\%, respectively. These modes are not visible when coupling with the tapered fiber alone. (d) Lorentzian fit to a high-Q mode at short wavelengths, showing a Q of $\sim 134,000$. (e)  "Shark-fin" shaped transmission dip characteristic of optical nonlinearities when pumping at ${\sim 138}{\mu \mathrm{W}}$ of measured power.}
\end{figure*}

In this study, we investigate the optical properties of a suspended silicon resonator etched via angled-etching (Fig. \ref{Fig2}(b)). As reported elsewhere\cite{Burek2014}, by selectively widening the patterned area and precisely timing the etch, we can create a fully suspended resonator with supporting sections ensuring sufficient distance ($\sim 2 \mathrm{\mu m}$) from the substrate. The nominal width of the resonator is ${1}\mathrm{\mu m}$, while the support region is ${1.1}\mathrm{\mu m}$ wide. The bending radius of the resonator is ${25}\mathrm{\mu m}$ and the etching angle is seen from SEM to be $\sim 65 ^\circ$. The cross-sectional area is sufficiently large to support several transverse modes in the structure. Fig. \ref{Fig2}(a) shows the mode profile of two such modes simulated in COMSOL. The top mode is the fundamental mode for the structure, with a calculated effective index of $3.19$. The higher-order mode (bottom) shows a calculated effective index of $1.52$, significantly lower than that of the fundamental. Seen through the lens of coupled mode theory, this implies that the higher-order mode should couple more easily to the bare optical fiber (effective index $\sim 1.4$). Furthermore, as seen in the mode profiles, the higher-order mode has more electric field concentrated on the sides of the waveguide. This implies that it interacts more with any surface roughness and adsorbed molecules, limiting the Q-factor of this mode family. In contrast, the fundamental mode is relatively isolated from the surface, resulting in a higher predicted Q-factor. In addition to these considerations, the multimode nature of the silicon resonator may increase losses during the transition region by the supports, causing coupling between mode families and limiting the observed Q-factors. For these reasons we do not necessarily expect our Q-factors to be limited by material absorption. Crucially, however, we anticipate that the fundamental mode has a much larger Q-factor than any other mode.

After loading the fiber with the silicon waveguide, the fiber was brought near the micro-resonator. The transmission spectrum was monitored continuously over a narrow bandwidth as the fiber was moved closer. The position of the fiber was controlled in ${50}\mathrm{nm}$ increments via stepper motor. Because the effective index of the loaded waveguide changes along its direction of propagation, all three spatial dimensions were used to tune the coupling to the resonator. Within the context of coupled mode theory, the coupling constant ($\kappa$) was tuned by changing the height of the fiber off the substrate and its lateral distance to the resonator, while the effective index contrast ($\delta$) was modified by translating the fiber along its length. 


The data shown in Fig. \ref{Fig3} were taken at optimized coupling locations either at a bare fiber section (red) or at the section containing the loaded waveguide (blue). A tunable telecom laser (Santec TSL-510) scanned the resonator as the transmission collected by a photoreceiver was monitored. Under a large scan range, the difference between the loaded and unloaded section is apparent when looking at shorter wavelengths (Fig \ref{Fig3}(a)). The increased coupling bandwidth is a result of the better index-matching. 

In the studied geometry, the length over which there is significant mode overlap between the resonator and the fiber is fixed due to the curvature of the resonator. Furthermore, the effective indices of the silicon device and resonator both vary strongly as a function of wavelength, whereas the index of the bare fiber varies weakly so. Because the loader waveguide and resonator have similar cross-sections, we can expect that their indices have a similar dependence on wavelength. In the context of coupled-mode theory, this implies that $\delta$ for the waveguide-loaded fiber-resonator system depends weakly on wavelength whereas $\delta$ for the bare fiber-resonator system depends on it strongly. This, in turn, imparts a strong wavelength-dependence on the coupling into high-index resonators from bare fibers (Fig. \ref{Fig3}(b)).

In addition to increased coupling bandwidth, transmission measurements of the resonator with the waveguide-loaded fiber showed the ability to access the high-effective index, fundamental modes of the device. Fig \ref{Fig3}(c) shows a scan taken at an optimized coupling position for wavelengths around ${1520}\mathrm{nm}$. High-Q resonances can be seen at $\sim{1518.4}\mathrm{nm}$ and $\sim{1521.5}\mathrm{nm}$. The group index can be calculated from the free spectral range (FSR, $\Delta \nu$) as $\Delta \nu = c/(n_g L)$ where $L$ is the resonator path length. These two modes give $n_g \approx 3.6$, which is close to the calculated value of $n_g = 3.85$. The discrepancy can be explained by differences between the simulated and actual waveguide dimensions. The modes are high-Q ($\sim 130,000$ and $\sim 40,000$, respectively), suggesting that they are relatively well-isolated from any surface scattering or absorption compared to higher-order modes (Q $\sim \ 10^4$). Furthermore, both show large transmission contrast ($\sim 60\%, \ \sim 80\%$, respectively), demonstrating that the waveguide-loaded fiber can efficiently transfer energy to the high-index modes of the resonator. The lower Q-factor of the second mode may be due to coupling with other resonances in the system\cite{Ramelow2014}.

To further demonstrate the ability to transfer large amounts of power to the resonator, we optimized the coupling at a particular wavelength and then increased the power of our laser until we observed optical bistability\cite{Xu2006}. As the laser is tuned across the resonance, the resonance peak is red-shifted until some critical detuning where the resonance transitions to a regime where it is no longer coupled (Fig. \ref{Fig3}(e)), resulting in a characteristic "shark-fin" shape. Additionally, no degradation in coupling was observed for higher powers, up to the maximum laser power available.

In summary, we have demonstrated a versatile technique which enables coupling from an optical fiber to a free-standing high-index micro-resonator with peak efficiency approaching 80\%. This result enables the rapid and large-scale optical probing of material systems useful in nonlinear and quantum optics such as diamond\cite{Burek2014}, chalcogenides\cite{Broaddus2009a}, lithium niobate\cite{Wang2014}, or III-Vs\cite{Mitchell2013}. Shorter operating wavelengths for quantum photonics with single-photon emitters like NV centers in diamond micro-cavities can be obtained by fabricating the loader waveguide with high-index materials transparent at visible wavelengths\cite{Patel2015, Choy2012}. Furthermore, the device can be tailored to the specific resonator by modifying the geometry of the coupling region, leading to highly efficient coupling directly to optical fibers. 

%
%

%

\begin{acknowledgments}
This work was performed in part at the Center for Nanoscale Systems (CNS), a member of the National Nanotechnology Infrastructure Network (NNIN), which is supported by the National Science Foundation under NSF award no. ECS-0335765. CNS is part of Harvard University. This work was supported by the DARPA SCOUT program through grant number W31P4Q-15-1-0013 from AMRDEC. P. L. is supported by the National Science Foundation Graduate Research Fellowship under Grant No. DGE1144152. M. J. B. was supported in part by the Natural Science and Engineering Council (NSERC) of Canada and the Harvard Quanum Optics Center (HQOC).

\end{acknowledgments}

\bibliography{loaderbib}

\end{document}